\DocumentMetadata{}
\documentclass[sigconf,screen,nonacm]{acmart}

\AtBeginDocument{%
  \providecommand\BibTeX{{%
    \normalfont B\kern-0.5em{\scshape i\kern-0.25em b}\kern-0.8em\TeX}}}

\usepackage{subfigure}
\usepackage{multirow}
\usepackage{xcolor}

\usepackage{amsmath}
\usepackage{multicol}
\usepackage{listings}
\usepackage{wrapfig}
\usepackage{algorithm}
\usepackage{algpseudocode}

\definecolor{codegreen}{rgb}{0,0.6,0}
\definecolor{codegray}{rgb}{0.5,0.5,0.5}
\definecolor{codepurple}{rgb}{0.58,0,0.82}
\definecolor{backcolour}{rgb}{0.95,0.95,0.92}

\title[Decoupled Edge Physics algorithms for collaborative XR simulations]%
      {Decoupled Edge Physics algorithms for collaborative XR simulations}

\begin{document}

\author{George Kokiadis}
\orcid{0000-0002-0858-5440}
\affiliation{%
  \institution{FORTH - ICS, University of Crete, ORamaVR}
  \city{}
  \country{}
}

\author{Antonis Protopsaltis}
\orcid{0000-0002-5670-1151}
\affiliation{%
  \institution{University of Western Macedonia, ORamaVR}
  \city{}
  \country{}
}

\author{Michalis Morfiadakis}
\orcid{0009-0006-4737-6285}
\affiliation{%
  \institution{University of Crete, ORamaVR}
  \city{}
  \country{}
}

\author{Nick Lydatakis}
\orcid{0000-0001-8159-9956}
\affiliation{%
  \institution{FORTH - ICS, University of Crete, ORamaVR}
  \city{}
  \country{}
}

\author{George Papagiannakis}
\orcid{0000-0002-2977-9850}
\affiliation{%
  \institution{FORTH - ICS, University of Crete, ORamaVR}
  \city{}
  \country{}
}

\renewcommand{\shortauthors}{Kokiadis, Protopsaltis, Morfiadakis, Lydatakis, Papagiannakis}

\begin{abstract}
This work proposes a novel approach to transform any modern game engine pipeline, for optimized performance and enhanced user experiences in Extended Reality (XR) environments. Decoupling the physics engine from the game engine pipeline and using a client-server N-1 architecture creates a scalable solution, efficiently serving multiple graphics clients on Head-Mounted Displays (HMDs) with a single physics engine on edge-cloud infrastructure. This approach ensures better synchronization in multiplayer scenarios without introducing overhead in single-player experiences, maintaining session continuity despite changes in user participation. Relocating the Physics Engine to an edge or cloud node reduces strain on local hardware, dedicating more resources to high-quality rendering and unlocking the full potential of untethered HMDs. We present four algorithms that decouple the physics engine, increasing frame rates and Quality of Experience (QoE) in VR simulations, supporting advanced interactions, numerous physics objects, and multi-user sessions with over 100 concurrent users. Incorporating a Geometric Algebra interpolator reduces inter-calls between dissected parts, maintaining QoE and easing network stress. Experimental validation, with more than 100 concurrent users, 10,000 physics objects, and softbody simulations, confirms the technical viability of the proposed architecture, showcasing transformative capabilities for more immersive and collaborative XR applications without compromising performance.
\end{abstract} 

\maketitle


\section{Introduction}

Immersive collaborative XR experiences demand realistic simulations for optimal QoE \cite{protopsaltis2024virtual}. In mobile-XR environments, efficient physics algorithms are essential for handling 3D object animations, transformations, and soft-body deformations while maintaining interactivity. 
Maintaining a minimum frame rate of 60fps is essential for fluid XR experiences, alongside high-resolution, low-latency graphics rendering.
Although modern game engines tackle many physics simulation challenges effectively, specific material physics aspects, such as liquid or deformable surfaces in XR scenes, are often overlooked, leading to unsatisfactory rendering. Standalone XR headsets with limited processing power often simplify physics models, affecting behavior alignment with expected norms. Tethering to high-end workstations improves processing power but limits user mobility, disrupting immersion.

\begin{figure}[h]
  \includegraphics[width=8.5cm]{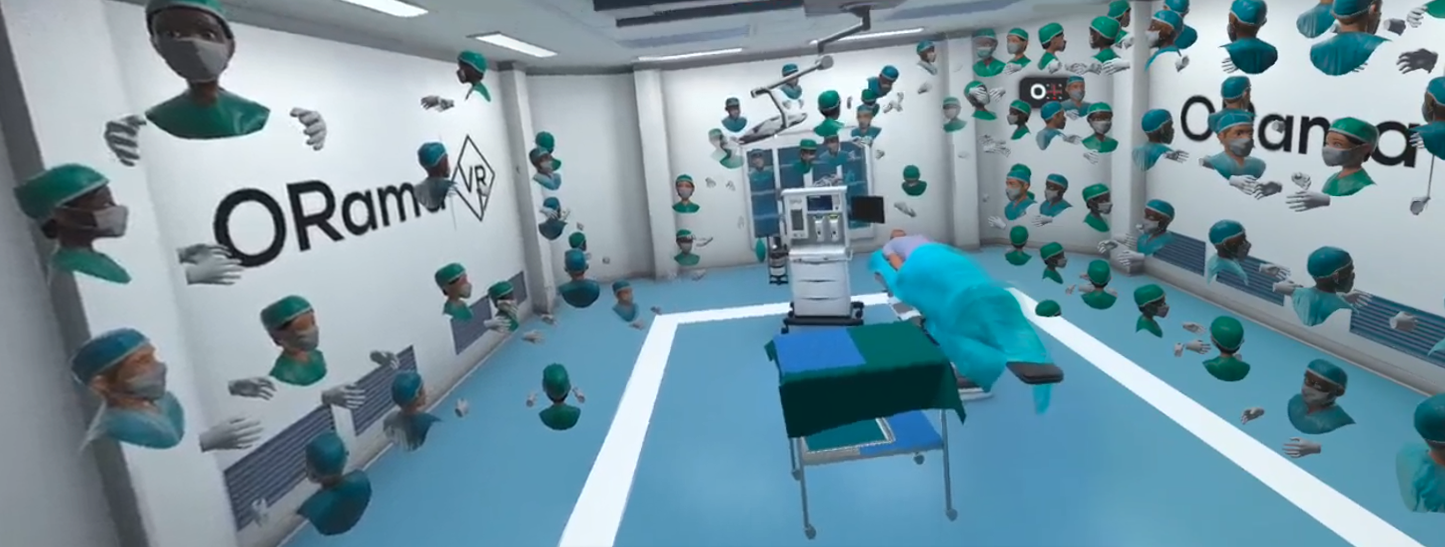}
\caption{Amphitheatrically placed 100 VR users around the patient}
\label{fig:PhysVisual} 
\end{figure}

In pursuit of advancing the efficiency and adaptability of modern game engines for untethered HMDs in immersive XR environments, this paper introduces a novel approach to dissect the native physics simulation engine from the main application. The primary goals of this decoupled physics unit are to minimize the total frame time for XR applications and to support real-time interactivity with multiple objects and enhance multi-player sessions, without imposing user limitations or QoE degradations. Through this approach, we seek to create a seamless and immersive XR gaming experience while addressing the challenges posed by the limitations of untethered HMDs.

\section{Related Work}\label{sec:2}

Untethered XR HMDs face significant challenges due to their relatively low processing power, which complicates the real-time simulation of realistic XR scenes and interactions. Immersive XR experiences impose strict requirements on low latency computations for user interactions and high-fidelity rendering for virtual worlds. 

Previous works analyzed the game engine's architecture \cite{gregory2018game}, the inter-calls between its modules, the CPU/GPU consumption, and resource requirements \cite{messaoudi2015dissecting}. Furthermore studies have assessed the computational load and power consumption on client devices \cite{nyamtiga2022edge}, the trade-offs between video quality and latency \cite{mehrabi2021multi}, and offloading rendering tasks \cite{lai2017furion, messaoudi2017using}. In that respect, micro-services \cite{newman2021building} were explored in mobile cloud integration and back-end architectures for MMORPGs \cite{liu2018integrating, vaha2017applying}. The offloading of modern game engine processes to cloud-edge necessitates considering their adaptability as service-oriented architecture or micro-service architecture, addressing the monolithic architecture challenges \cite{newman2019monolith, makris2021cloud, makris2023streamlining, messaoudi2018toward, zhou2023computing}.

Remote rendering architectures were also utilized to alleviate the computational burden on untethered XR HMDs, that typically involve a monolithic cloud service responsible for performing all rendering, game logic and physics computations, with the encoded video streamed to the lightweight HMD via high-speed networks. Notable solutions include NVIDIA's CloudXR
\footnote{\url{www.nvidia.com/en-us/design-visualization/solutions/cloud-xr/}} and the open source ALVR\footnote{\url{github.com/alvr-org/ALVR}}. While monolithic remote rendering architectures have shown promising, their the data-intensive nature imposes substantial network and edge-cloud infrastructure requirements. Each HMD user necessitates a corresponding GPU-enabled workstation on the edge or cloud, amplifying resource demands, especially in multi-user gaming scenarios. To accommodate such significant network challenges, often requires leveraging high-speed 5G/6G networks, that minimize latency and increase the available bandwidth.

Considering these challenges, there's growing interest in dissecting modern game engines and utilize a distributed pipeline. As interactive XR scenes involve intense physics computations, offloading the physics engine as a edge-cloud service is a promising solution. In that respect, edge-physics frameworks as in \cite{friston2021quality} propose streaming at the scene-graph level, reporting significantly lower computational overhead, bandwidth, and latency compared to video streaming. Another approach \cite{kurt2023edge} involves utilizing the open source Bevy game engine in conjunction with a remote dedicated physics server, such as Rapier, to simulate non-XR scenarios of simple scenes without real-time user interactions. Moreover, established physics engines, like NVIDIA PhysX and Havok, offer support for distributed physics simulation, allowing developers to distribute physics computations across multiple machines. These alternative solutions often introduce great complexity in the development process of an XR solution, as they are not generalized, so this involves hard-coded scene specific details.

Decoupling the physics engine from monolithic game engines like Unity and Unreal Engine poses a formidable challenges. The tight integration of the physics engine with core subsystems such as rendering, input handling, and game logic and the numerous inter-calls between them complicates the separation process. Synchronization of the simulation state in real-time for multi-user interactive scenarios is a complex task, that imposes strict requirements for QoE, especially under degraded network conditions. Achieving physics engine decoupling requires careful consideration and potentially extensive modifications to the engine's internals.

\section{Goals and Constraints}\label{sec:3}

In contemporary computational architectures, particularly within the realm of game engines and XR systems, a monolithic design is commonly employed, wherein both physics and graphics computations are conducted on the same hardware. This conventional approach can often impose significant computational burdens on devices, especially on XR HMDs.

This research work aims to decouple physics computations from any modern game engine to revolutionize immersive experiences, particularly for untethered HMDs, by addressing several critical goals while adhering to constraints inherent in distributed XR pipelines. By offloading physics computations to edge or cloud nodes, the final system endeavors to achieve optimal QoE to maintain user immersion, even during interactions with softbodies or scenes with a high number of objects, which traditionally strain the processing power of standalone devices. This research not only enables the realization of complex simulations but also scales the virtual environment to accommodate a vast number of physics objects, enriching the depth and interactivity of immersive scenes. This optimization requires low application latency, facilitated by low network latency and high-bandwidth networks to ensure seamless interaction and rendering of complex virtual environments.

The decoupling strategy serves a dual purpose of reducing the CPU load on untethered XR HMDs and enhancing performance in scenarios involving intensive physics simulations. By alleviating the computational burden on onboard processors, the research aims to not only boost frame rates but also to elevate the quality of experience for users, fostering smoother visuals and more responsive interactions. Moreover, this distributed XR pipeline aims to  offer tangible benefits such as increased battery life and enhanced user mobility, thereby ensuring prolonged and uninterrupted immersive experiences. An overloaded CPU acts as a bottleneck\footnote{\url{https://www.intel.com/content/www/us/en/gaming/resources/what-is-bottlenecking-my-pc.html}} in the overall system performance, even if the GPU is capable of handling more tasks. Offloading physics computations from the primary device reduces the computational load on HMDs, optimizing the overall pipeline and ultimately the utilization of the GPU, allowing it to operate at full capacity and improving the overall graphical performance. Additionally, segregating physics computations allows for more sophisticated simulations, such as soft body dynamics and managing a large number of physics objects, without adversely impacting the frame rate, as these computations can run on a dedicated server or processor.

A high frame rate is crucial in XR environments, as low frames per second (FPS) can lead to motion sickness and break the user's sense of immersion. By ensuring that the physics calculations do not interfere with the graphical frame rate, users are less likely to experience nausea and other XR-induced discomforts.

The Metaverse has made the requirement for dynamic and robust multi-user and interactive experiences in XR environments obvious. In this research we aim to  maintain the XR session's physics state independently from the users' devices, to support up to 100 collaborating interactive MR users in the same scene (see figure \ref{fig:PhysVisual}). To enrich the gameplay experience and enhance the realism and immersiveness of the environment, the distributed pipeline will allow users to real-time interact with objects, in ways that mimic real-world interactions, such as grabbing objects or interacting with softbodies. Synchronization is paramount for user interactions and multi-user sessions, enabling seamless collaboration and dynamic interactions within XR environments. Network usage should remain acceptable for average home/business networks to ensure widespread adoption and accessibility.

Compatibility with various modern game engines is essential to ensure the versatility and accessibility of the designed solution. In that respect, the developer experience must be streamlined, with minimal hindrance and seamless integration of the decoupled physics component. The distributed XR pipeline must be adaptable to the game engine development cycle, ensuring seamless integration of the decoupled physics engine in the game development workflow.

Ultimately, the research endeavors to leverage edge-cloud infrastructure to offload all physics computations, transcending the limitations of standalone XR HMDs and enabling the realization of realistic operations that include heavy physics simulations. 
Overall, by meeting these goals and adhering to the constraints of distributed XR pipelines, this research work seeks to advance the capabilities and accessibility of XR applications, enabling richer, more immersive, and collaborative virtual experiences for users. 

\section{Altering the Game Engine pipeline}
\label{sec:4}
The Entity-Component-System (ECS) pattern \cite{nystrom2014game} has been widely adopted by modern game engines to facilitate the construction of complex game systems, empowering diverse and immersive interactive experiences while maintaining flexibility and scalability. Modern game engines also utilize the scene graph, a hierarchical structure that organizes entities and components. Several modern game engines, including Unity's DOTS (Data-Oriented Technology Stack), Unreal Engine's ECS, Bevy\footnote{\url{https://bevyengine.org/}}, and the open-source Godot Engine\footnote{\url{https://godotengine.org/}}, have embraced the ECS pattern to streamline game development and enhance performance. Most of these engines tightly integrate a physics engine into their core functionality under the hood, providing accurate and realistic physics simulations for a wide range of applications. Specifically, Unity and Unreal Engine utilize NVIDIA's PhysX physics engine, Bevy cooperates with the Rapier\footnote{\url{https://rapier.rs/}} physics engine, and Godot uses the open-source Bullet physics engine\footnote{\url{https://github.com/bulletphysics}}.

\subsection{Multiple Users and N-1 Architecture}
\label{ssec:multiuser_architecture}

In multi-user virtual environments, each instance of a Multi-user Graphics Engine application traditionally operates its own dedicated physics engine. However, an innovative modification to this setup involves implementing a centralized physics engine micro-service. This micro-service, hosted on a server within the cloud-edge continuum, can be utilized in a 1-N relationship by multiple graphics applications engaged in the same game session. Consequently, this centralized model allows the physics micro-service to perform simulations and then stream the results to all connected graphics rendering applications (see figure \ref{fig:PhysVisual}). This architecture not only facilitates the collaboration of a significant number of concurrent users but also enhances the efficiency and scalability of resource utilization.

\begin{figure}
  \includegraphics[width=8cm]{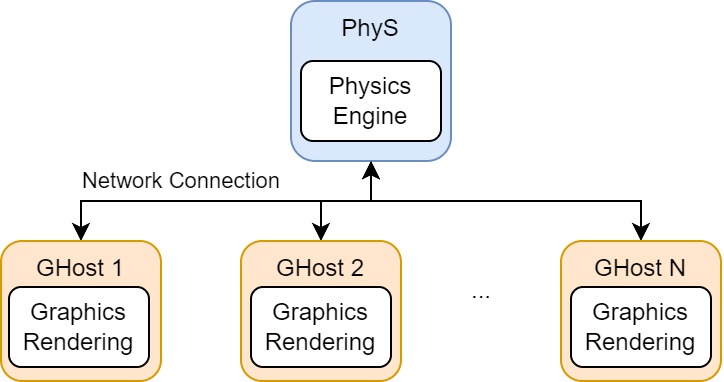}
  \includegraphics[width=8cm]{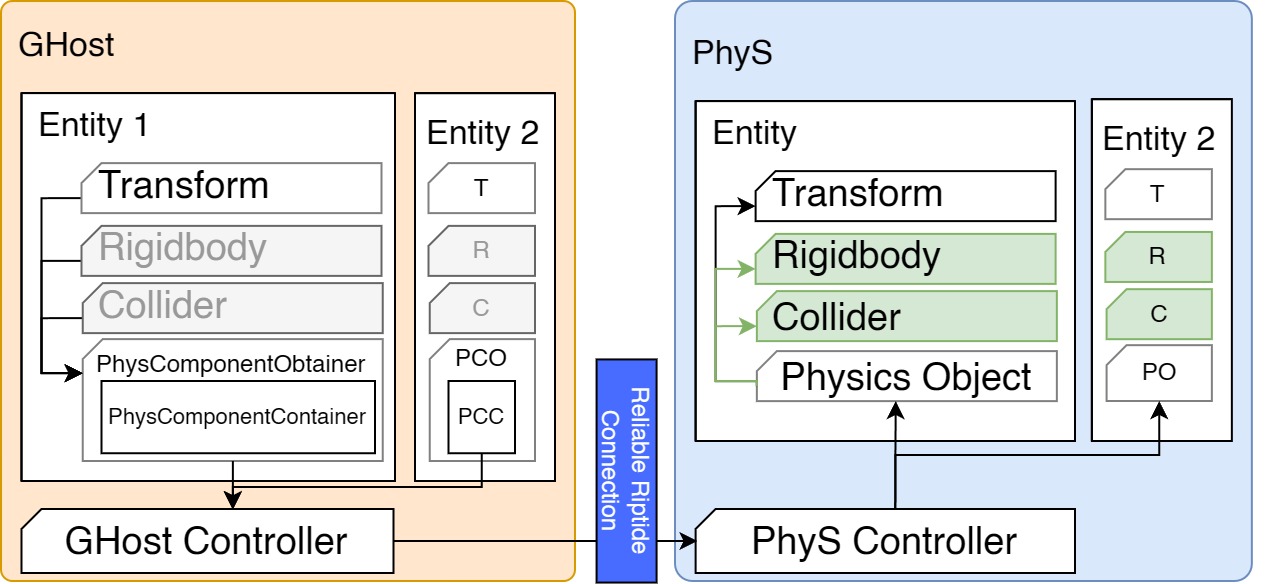}
\caption{Top: Overview of the N-1 Decoupled approach. Bottom: Process of dissecting an entity}
\label{fig:simple_architecture_go_dissection} 
\end{figure}

The modification of a game engine pipeline delineates two key units with bidirectional communication capabilities: the Graphics Host (GHost) and the Physics Server (PhyS) (see figure \ref{fig:simple_architecture_go_dissection}). The GHost unit encompasses the entirety of the game engine graphics pipeline and operates without any active Physics Components during game play. GHost is responsible for maintaining the game logic and performing the rendering. 

Conversely, the PhyS unit solely manages the entirety of physics computations on behalf of the game engine. The PhyS unit runs in optimal Headless mode, where all processing, communication and calculations are performed without the rendering pipeline, to be as lightweight as possible.

The primary objective of this distributed game engine pipeline is to enable every Entity within the scene to be fully simulated by the PhyS, thereby alleviating the computational load on the untethered HMD operated by the GHost. To ensure coherence and real-time performance between these two units, network communication and synchronization is achieved through Riptide networking\footnote{\url{https://github.com/RiptideNetworking/Riptide}}. The lightweight and open-source nature of this networking library allows the exchange of only the absolutely necessary messages, with minimal overhead or added processing from Riptide itself. We predominantly utilize Riptide's "unreliable" connection type, which, akin to UDP, is strategic for minimizing network latency. This type of connection is suitable for streaming applications where occasional packet loss is tolerable because subsequent messages quickly follow. However, for crucial communications that must be reliably delivered - such as object initialization or significant Entity state updates, or object deletion — we use Riptide's "reliable" connection types to ensure these important packets reach their destination.

\subsection{Dissecting an Entity}
\label{ssec:dissecting_gameobject}
The scene graph is the fundamental building block of a game scene, representing a hierarchy of entities in the game world, that serve as containers for components that define its properties, behavior, and appearance, ultimately contributing to the overall interactivity and visual representation of the game. Entities can have a variety of components attached to them, each serving a specific purpose. An entity can interact with the physics engine through the use of certain components and features that allow the simulation of physical behaviors.

The development of a game application with decoupled physics behavior from an entity involves the usual entity creation process in GHost with an additional "PhysComponentObtainer" (PCO) component. This component is tasked with extracting and packaging all physics-related scripts from the respective entity, along with the object's current transformation data, into a data structure named "Phys Component Container" (PCC). The PCC holds all the necessary information required to initialize the respective entity in the PhyS. When all physics-related information has been extracted from the Entity, the components are removed from the specific entity, to prevent Physics simulations from being computed on the GHost unit. After this initialization process is complete, in the GHost, each PCO component creates a "Graphics Object" (GrO) Component, attaches it to its Entity, and then destroys itself.

The initialization process also ensures that all entities in the GHost scene graph, containing physics components, are hierarchically replicated in the PhyS with only its physics components present (see Algorithm \ref{alg:selective_transmission}). This requires any parent that may not contain physics components to still be present in the PhyS. In this case, a PCO must also be placed in the parent, which will generate a PCC which only contains transform information.

\begin{algorithm}
\caption{Selective replication of entity and its parent based on physics components}
\label{alg:selective_transmission}
\begin{algorithmic}[1]
\For{each Entity in GHost scene graph}
    \If{Entity has physics components}
        \State Send Entity to PhyS
        \If{Parent of Entity exists}
            \State Send Parent Entity to PhyS
        \EndIf
    \EndIf
\EndFor
\end{algorithmic}
\end{algorithm}

The system also integrates both a "GHost Controller" (GHC) and a "PhyS Controller" (PhC). Each controller is responsible for managing its respective domain: the GHC oversees all Graphics Objects, while the PhC handles all Physics Objects in the application's scene. These components are also responsible for exchanging messages with each other, to allow for the synchronization of the scene graph in both units. During the initialization stage, the GHC is aware of all PCO Components, gathers all PCCs from the respective PCOs, and transmits them to the PhC. (see Algorithms \ref{alg:go_dissection1}, \ref{alg:go_dissection2}, Figure 
\ref{fig:simple_architecture_go_dissection})

For each PCC received in the PhyS from the GHost, a new Entity is created. In this new Entity, a "Physics Object" (PO) Component is initialized, and the respective PCC is passed to it. The PO is responsible for doing the reverse operation that the PCO does, creating all Physics Components included in the PCC. As a result, we effectively create two distinct representations of each entity in the scene graph: the \textit{Graphics object} in the GHost and the \textit{Physics object}  on the PhyS. This dual representation ensures unmodified development process for developers while seamlessly integrating with the advanced simulation capabilities of the physics engine.

\begin{algorithm}
\caption{Initiation of GHost entity dissection}
\begin{algorithmic}[1]
\For{all Entity in Scene graph with physics comp.}
    \State Initialize PhysComponentContainer
    \For{all physics-related comp. in Entity}
        \State Get all Parameters of the Component
        \State Store Parameters in PhysComponentContainer
        \State Remove Component from Entity
    \EndFor
    \State Assign unique EntityID to PhysComponentContainer
    \State Copy PhysComponentContainer to ContainerList in GHC
    \State Delete PhysComponentContainer component
    \State Initialize Graphics Object Component with EntityID
\EndFor
\State Send all containers in the ContainerList to PhyS
\end{algorithmic}
\label{alg:go_dissection1}
\end{algorithm}

During the initialization of this process, a unique Entity ID is assigned to each Graphics and Physics object, which is consistent between them. This unique ID ensures that each GrO and its corresponding PO can synchronize accurately, maintaining a consistent state across both systems throughout the game play.

\begin{algorithm}
\caption{Initiation of replicated PhyS entity}
\begin{algorithmic}[1]
\For{all received PhysComponentContainer}
    \State Create a new Entity
    \State Initialize PO comp. with EntityID
    \For{all Physics comp. in PhysComponentContainer}
        \State Initialize Physics comp.
    \EndFor
    \State Retain scene graph hierarchy by assigning each object to their parent
\EndFor
\end{algorithmic}
\label{alg:go_dissection2}
\end{algorithm}
After the initialization process, the GHost has no physics components to simulate, leaving all physics simulations to the PhyS. The GrO's transformations are controlled by the PhyS, allowing the GHost to focus on accurately placing the rendered entities in their correct positions.

In specific instances, the GHost can send a "MoveToTransform" command to the PhyS to manually adjust an object's transform. However, these transformations are still subject to the laws of physics within the simulation. For example, if a user tries to move an object through another, the objects will collide instead of overlapping, due to their physical properties. The resulting physics simulated transform is what the GHost receives and presents in the HMD to the user.

\subsection{Multi-user Session initiation \& runtime}
\label{ssec:multiuser}
In our game environment, the PhyS micro service can be deployed either by one of the GHosts or through an automated cloud service provider. Once deployed, each GHost connects to the server using its IP and port, establishing a Riptide connection.

In a multi-user session, GHost applications, served by the same PhyS, will contain an identical scene graph. During the initialization phase of the Dissection, the GHost scene graph structure is replicated and transmitted to the PhyS (see figure \ref{fig:initialization_handcolliders-Bunny}). 

The PhyS unit is initialized by the first GHost unit in the multi-user session. Its scene graph is not hard-coded, but replicated from the first GHost's scene graph. This functionality allows compatibility with any GHost unit, promoting versatility and adaptability across different use cases and applications. Subsequent GHosts joining the session, immediately start their session participation by receiving entities' transform updates. 

\subsection{User Avatars}
\label{ssec:avatars}
In XR Multi-User environments, user avatars are essential for representing players within the digital space. Each avatar consists of several Entities, each designated by a unique identifier—combining a specific EntityID with the respective PlayerID from the session. This identification system ensures that each avatar's interactions and movements are distinctly linked to the corresponding player, enabling precise control over individual avatars in a multi-user setting.

Control over avatars and their movements traditionally comes through input devices like controllers and HMDs. The input transformations received from these devices are processed in the GHost and then communicated to the PhyS, through the transmission of MoveToTransform commands. The outcome of these commands is different for each case.

The hands interactor of the user's avatar are subject to physics. This means that if a user moves his hand into a space where it collides with an interactable object, a collision will occur. The real position of the user's hand in physical space may differ from the respective position in digital space due to these collisions. Should a user try to move their hand through an object, the hand and object will collide instead of overlapping, due to their physical properties.

The user's avatar follows the scene Camera, where the user's point of view is located. The avatar itself does not undergo physics simulations, since this would lead to interfering with an XR user's point of view, which can cause nausea, discomfort and disorientation. For this reason it only contains a kinematic rigid body component, that allows it to be moved in digital space without being subject to physical forces. While avatars do not have collider components, they must still be updated in the PhyS to ensure all players see consistent avatar positions.

\subsection{Collision Events \& Interaction}
\label{ssec:col_events}

Collision events are integral to fostering interactive game play and crafting engaging narratives within the game environment. Since the GHost has no way of detecting collisions, since no Physics Components exist, all collision detection occurs in the PhyS. When an Entity collides with another in the PhyS, the PO Component collects the EntityID of the collided Entity, and uses the PhC to send it is to the GHost where the GrO Component is updated with it. This allows the GHost to keep track of which entities are colliding with each other. This is important for implementing game logic, interactions and an overall responsive environment. 
\begin{figure}[h]
  \centering
  \includegraphics[width=8cm]{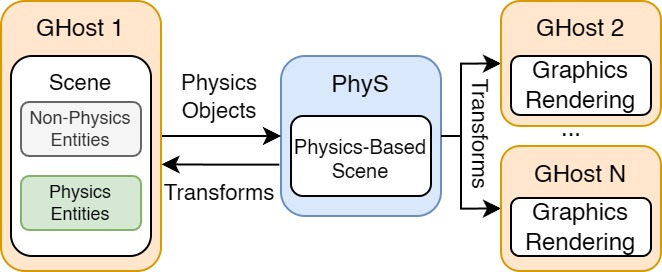}
  \includegraphics[width=4cm]{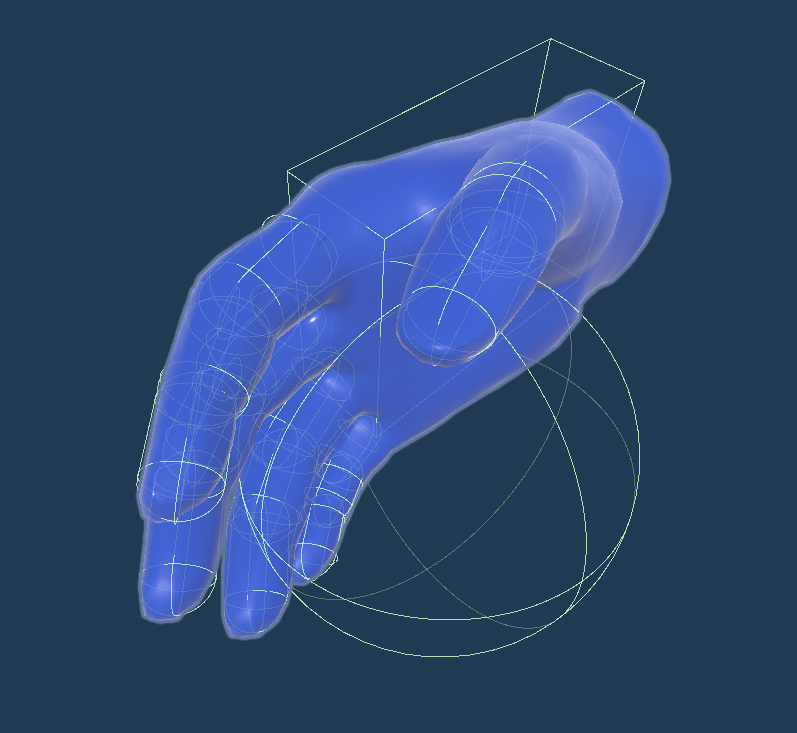}
  \includegraphics[width=4cm]{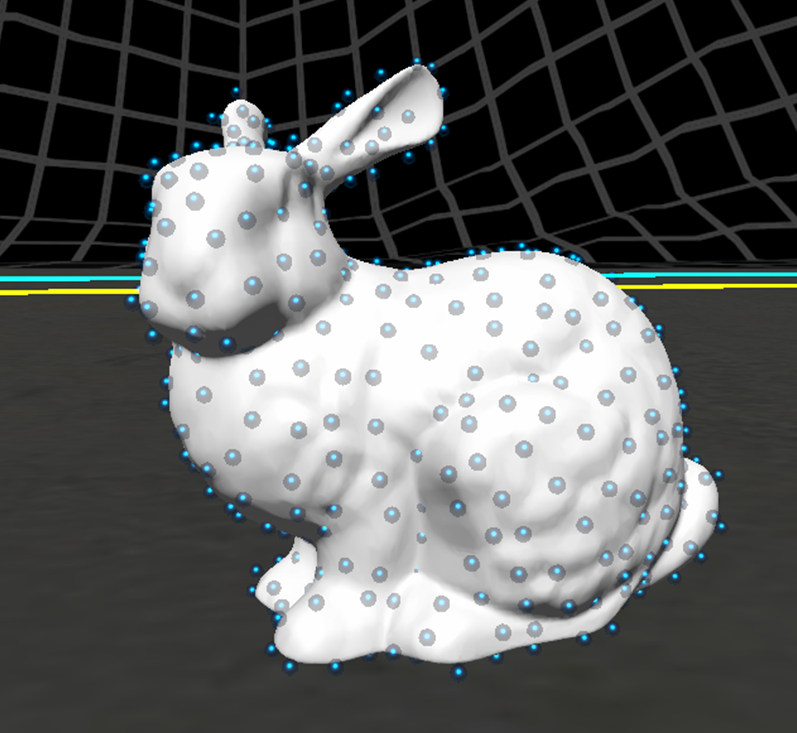}
\caption{Top: PhyS Initialization process. Bot-Left: Hand Colliders. Sphere Collider bellow the hand is the Trigger Collider. Bot-Right: Soft Body Bunny with particles and spring connections between them}
\label{fig:initialization_handcolliders-Bunny}
\end{figure}

To enhance the system's functionality, a new event-handling system was implemented in GHost, that handles "CollisionEnter" and "CollisionExit" events. These custom events are triggered accordingly, allowing for specific responses when Entities begin or end their interaction through a collision. This method of handling collision events allows the design of complex scenarios and story-driven game play, as it integrates dynamic events and interactions seamlessly into the narrative.

Collision events are fundamental for the interaction system within the game environment, particularly in how they facilitate object manipulation by the user. Every interactable entity, contains an "Interactable" component in the GHost. Each of the user's hands entity in the GHost contains an "Interactor" Component and Trigger Collider Components in the PhyS (see figure \ref{fig:initialization_handcolliders-Bunny}). When the user presses the "grab" button on their controller, the Interactor Component accesses the GrO Component, to acquire the Entity that the Interactor is currently hovering, as reported by the PO to the GrO. If the Entity contains an Interactable component, interaction starts. The interaction ends when the user releases the the "grab" button (see Algorithm \ref{alg:interactionsys}).
\begin{algorithm}
\caption{Interaction System}\label{alg:interaction}
\begin{algorithmic}[1]
    \While{grab button is pressed}
        \If{hovered entity has interactable component}
            \State Start interaction with entity
            \State Continue interaction while button is held
        \EndIf
    \EndWhile
    \State End interaction when button is released
\end{algorithmic}
\label{alg:interactionsys}
\end{algorithm}
Once an object is successfully grabbed, GHost performs calculations to determine the correct position the interactable entity should have, in relation to the user's hands. The position of the Interactable must be such that it appears within the user's hand. These computations, performed in the GHost at every frame, are required for realistic interactions.  The resulting position is sent to the PhyS using a MoveToTransform command. Similarly to how the movement of the hands is achieved, the position of the interactable is with respect to physics, so if a user tries to move an object through another, the objects will collide and not overlap.

\subsection{Simulating Softbodies}
\label{ssec:softbodies}
To simulate soft bodies in our system, we use a particle-based method \cite{kamarianakis2022progressive} where the mesh vertices of the model are clustered in particles (see figure \ref{fig:initialization_handcolliders-Bunny}). Each particle controls a set of vertices within a specific range, which allows for realistic deformation when forces are applied. Neighboring particles are interconnected with each other, forming a particle map, allowing force exertion in nearby particles. This approach ensures that movement and deformations are realistically portrayed while maintaining system performance.

This processing of soft bodies is handled within the PhyS. To achieve a coherent simulation, the positions of the particles are synchronized with the GHost. This synchronization ensures that both the physical interactions and the visual representations are consistent and accurate across the system, providing a seamless and realistic experience in the simulated environment.

\subsection{Relay server for local-physics compatibility}
\label{ssec:relay}

A relay server in collaborative XR environments acts as an intermediary that facilitates communication between XR clients. This server centralizes data traffic by receiving information from one client and redistributing it to others. This model is particularly beneficial in scenarios where direct peer-to-peer communication is impractical due to network constraints or when uniform data handling is critical for maintaining a cohesive virtual environment.

\begin{figure}[h]
  \centering{
  \includegraphics[width=8cm]{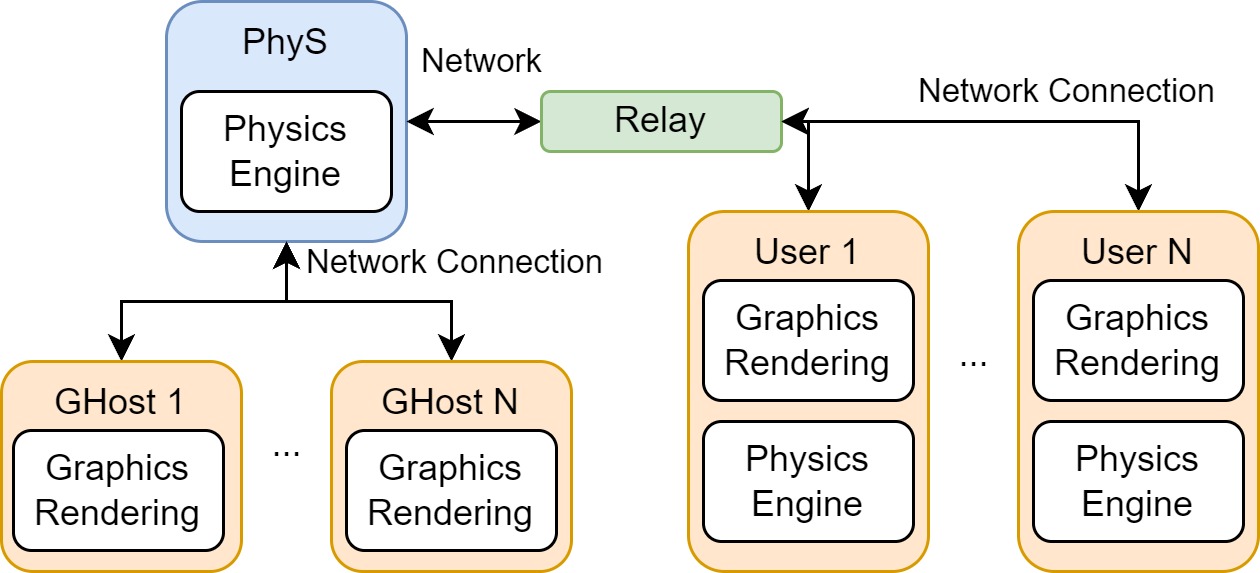}
  \includegraphics[width=8cm]{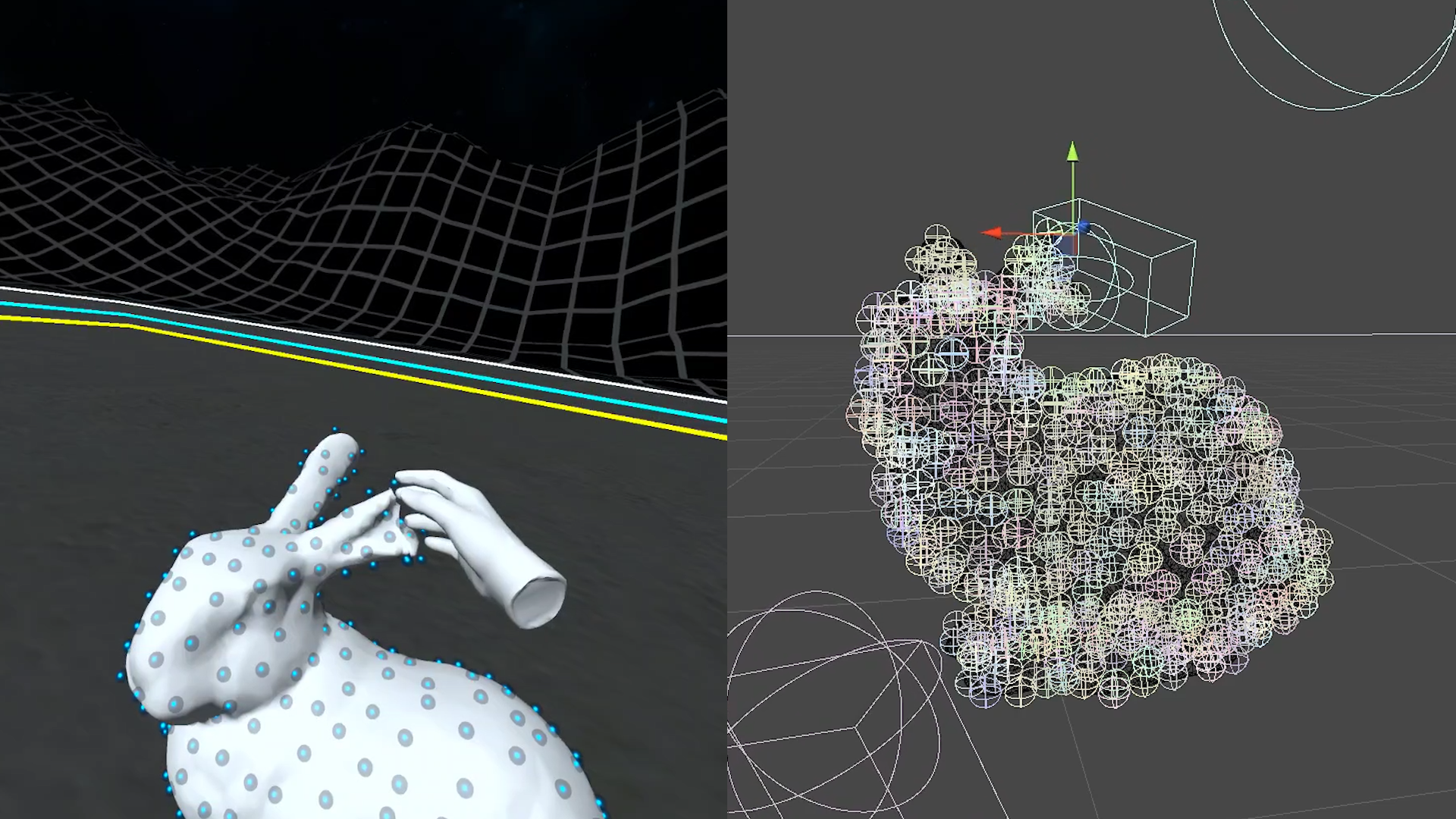}}
\caption{Top: Architecture with Relay Server and PhyS. Bottom: Softbody Experimentation Scenario.}
\label{fig:architecture_relay_ExpSoftbody}
\end{figure}

In our setup, the relay server and the physics server coexist to fit a diverse range of client capabilities and ensure a uniform experience across all participants in the session (see figure \ref{fig:architecture_relay_ExpSoftbody}). The physics server is available for clients that needs to offload intensive physics calculations, particularly beneficial for lower-end devices that might struggle with complex simulations. This server handles the heavy computational load, allowing these devices to maintain high performance without local resource strain. Meanwhile, the relay server maintains its role as the central coordinator for all client interactions, managing data synchronization. It ensures that updates, whether processed locally or computed by the physics server, are consistently distributed to all clients, keeping the virtual environment synchronized. This dual-server setup maximizes the efficiency of network and computational resources and provides flexibility and scalability, supporting a wide array of devices.

\subsection{Network optimisations for Relaying}
The described architecture involves two relaying servers. The PhyS, that besides physics simulations, relays transformation data to all GHosts in the session, and the relay server that we described in section \ref{ssec:relay}. Optimizing a relay server for sending involves minimizing the data sent across the network while ensuring game state consistency and responsiveness. In our solution we've developed strategies to improve how our relay server handles transformation data, ensuring our simulation stays fast and consistent.\\
\textbf{Selective Synchronization:} This method scrutinizes the positional and rotational states of XR objects since their last update, transmitting only those changes that exceed a set significance threshold. This ensures that only impactful alterations are communicated, conserving bandwidth by omitting minor updates.\\
\textbf{Bitmasking Strategy:} When changes are detected, we use a specific coding system to identify the kind of change—whether it's in position, rotation, or both. This method allows us to pack our data more efficiently, sending only the necessary data. This reduces the size and amount of data we send.\\
\textbf{State Update Mechanism:} On the other end, our system reads the incoming codes to figure out what has changed. It then updates only those parts of the local XR objects that need it, which keeps everything running smoothly without unnecessary work.\\
\textbf{Grouped updates:} The relay server transmits transform updates in groups rather than as separate messages for each entity. This grouping strategy reduces the cumulative overhead caused by the multiple header bytes included in each individual network message.\\
\textbf{Dynamic message size:} The size of the message the group is sent in is dynamic, meaning that only the absolutely necessary level of network usage is reached. By consolidating updates into one message, we significantly decrease the network load and enhance the efficiency of data transmission.\\
\textbf{Sending at Intervals:} The Relay Server employs a strategy of transmitting transform updates to the GHosts at fixed intervals for all Entities within the dissected environment. To further refine this approach, entities deemed as critical by the developer of the Application, such as interactable objects or user avatars, are updated at a higher frequency than non-critical items. This increased rate ensures a smoother experience by providing more frequent updates for Entities that significantly influence game play and user interaction. 

Despite the benefits of interval-based updating and batching, this method can introduce visual artifacts, such as stuttering or perceptible lags in the movement of synchronized objects. To address these issues and improve visual continuity, we have implemented a Dual Quaternion interpolator \cite{kamarianakis2023less, 
zikas2023mages}. This tool effectively smooths out the motion of objects between received updates, creating a more fluid and natural appearance. The interpolator calculates intermediate states by considering previous and current transform data, thus mitigating the impact of network-induced delays and providing a seamless visual experience.

\begin{figure}[h]
  \includegraphics[width=8cm] {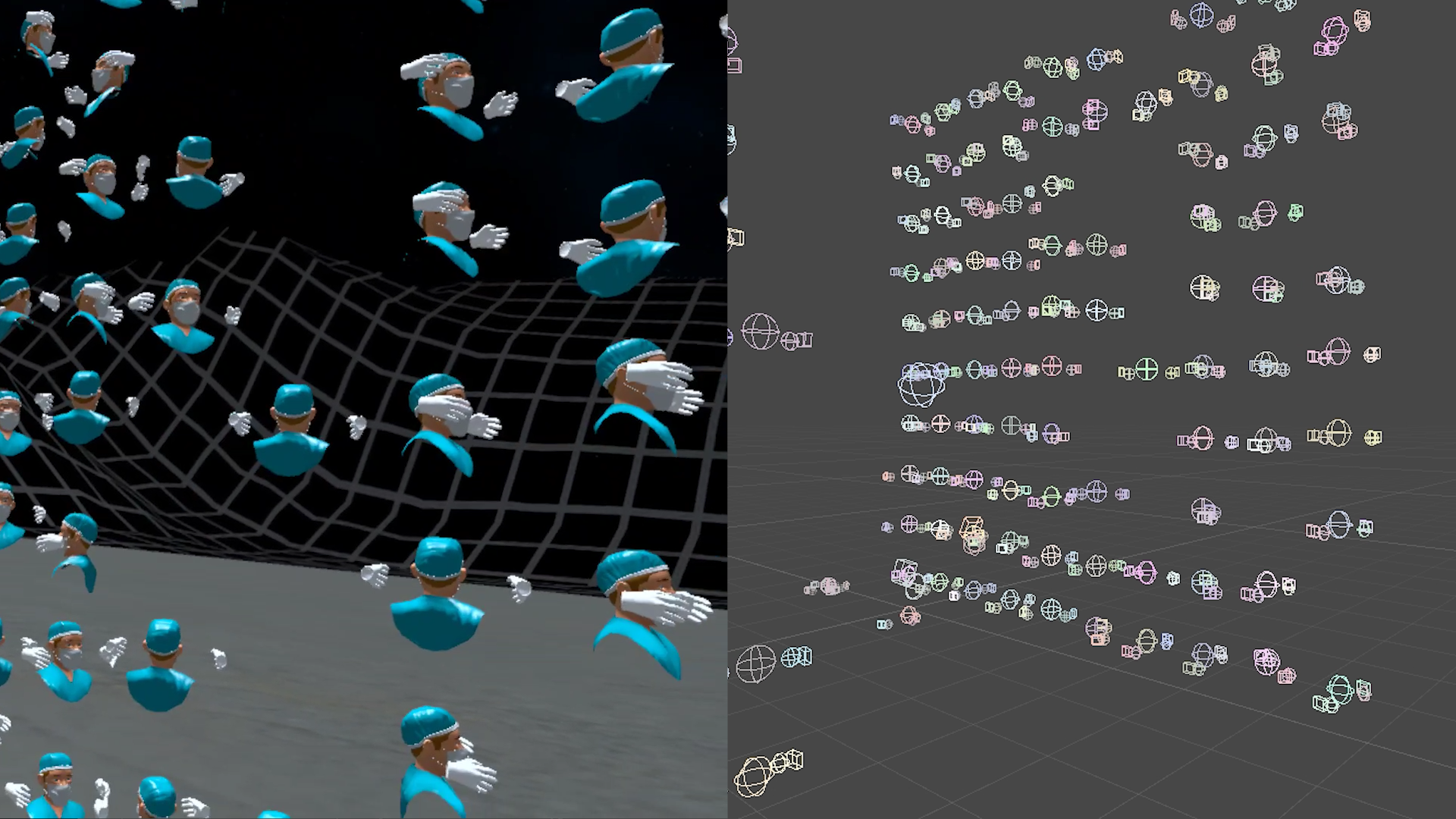}
  \includegraphics[width=8cm]{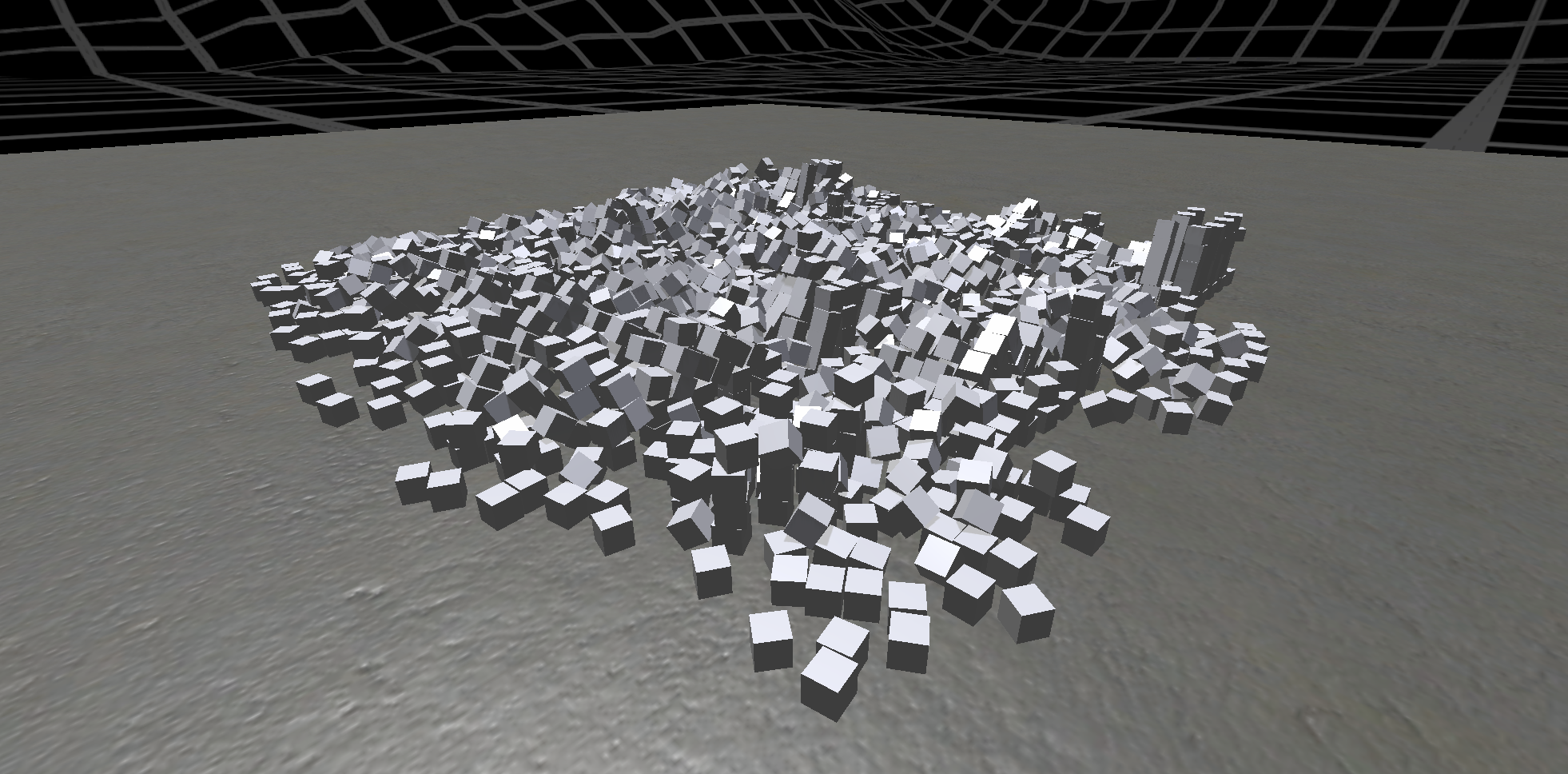}
\caption{Top: 100 CCU experimentation with HMD view and Physics objects in PhyS. Bottom: MultiObject Experimentation Scenario.}
\label{fig:ExpCCU_MultiObject} 
\end{figure}
This combination of strategic update intervals, batch processing of updates, and advanced interpolation techniques ensures that our network optimization efforts enhance the user's experience by creating a more responsive and engaging virtual environment.

\section{Experimentation}
\label{sec:5}
To accurately assess the device load and performance enhancements, our experiment will compare the effects of local versus decoupled physics processing. We will use a controlled setup with the HMD and a more powerful server running PhyS over a local area network. The goal is to determine if decoupled physics simulations improves performance and reduces load on the HMD. 
\begin{figure}[h]
  \includegraphics[width=8cm]{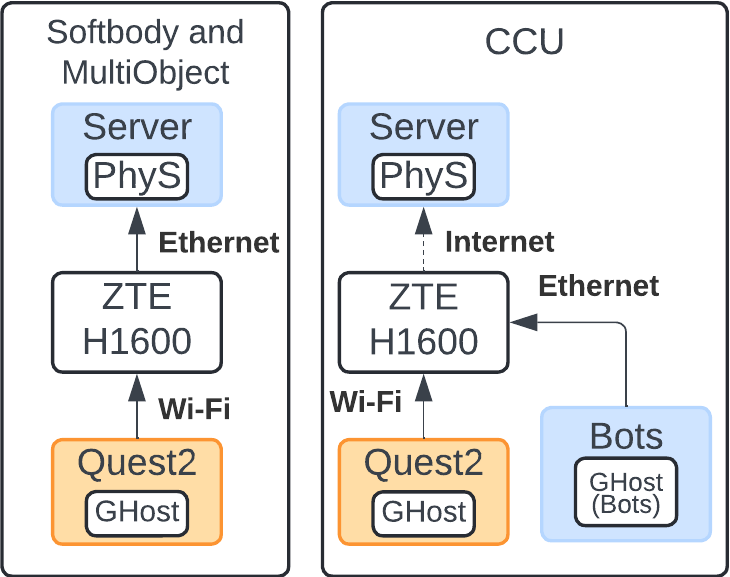}
  \includegraphics[width=8cm]{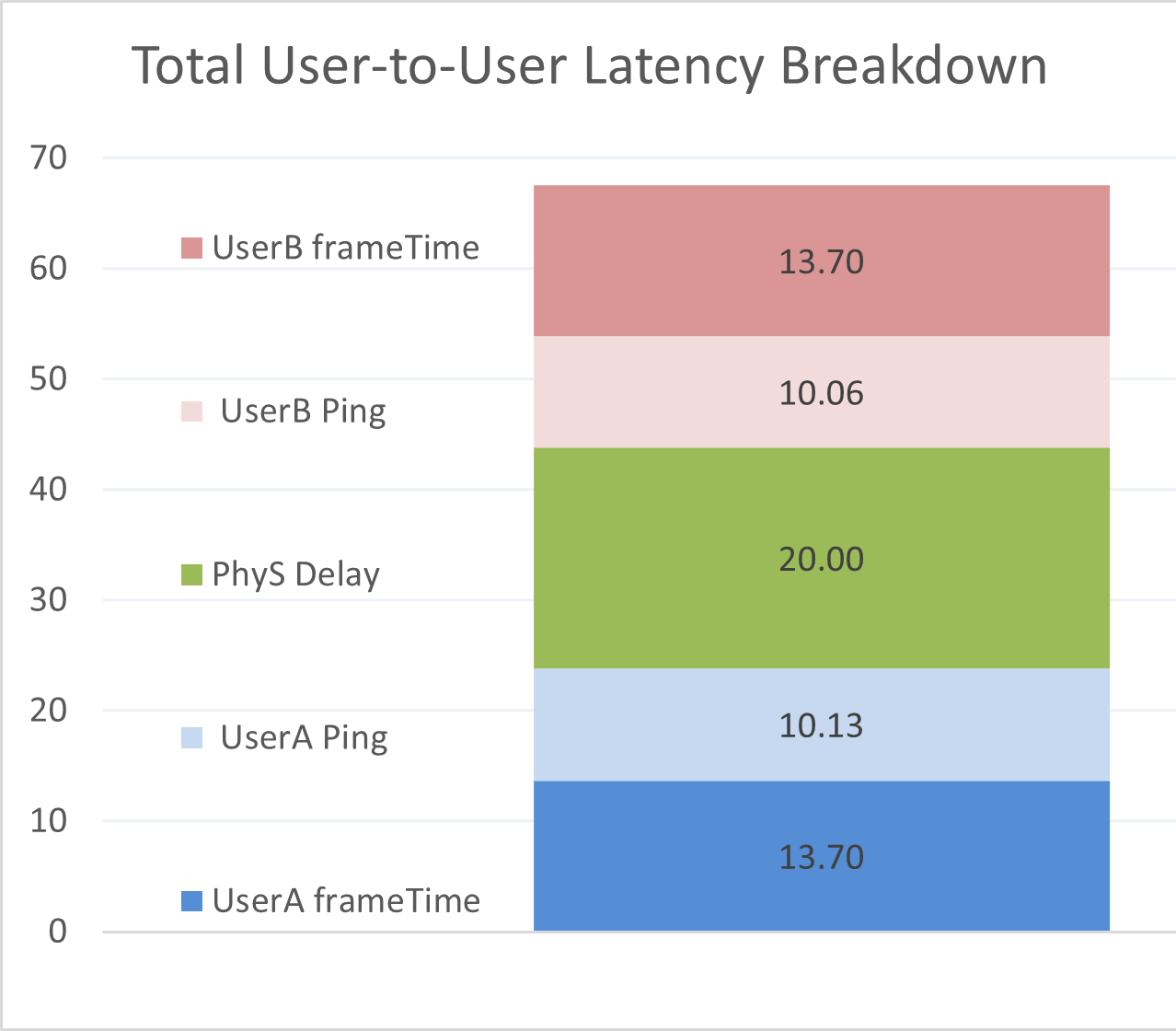}
\caption{Top: Experimental Setup. Bottom: Total user-to-user latency breakdown.}
\label{fig:ExpSetup} 
\end{figure}

Our experimental process (figure \ref{fig:ExpSetup}) includes three experimentation scenarios that will assess the distributed XR pipeline performance: a) the "Softbody" scenario deals with complex physics computations while the user interacts with varying numbers of soft body models (Bunny1 with 500 particles, Bunny3 with 1500 particles) (see figure \ref{fig:architecture_relay_ExpSoftbody}) b) the "MultiObject" scenario includes physics computations on a scene with varied the number of objects (500, 1000, 2000, 5000 and 10000) with active physics simulations (see figure \ref{fig:ExpCCU_MultiObject}), and c) the "CCU" (concurrent users) scenario that involves an interactive scene with multiple (100+) CCUs, gradually joining the XR session (see figure \ref{fig:ExpCCU_MultiObject}). This approach allows us to assess how offloading physics computations impacts user device performance across different levels of complexity and user load, while preserving acceptable QoE. Finally, we will experiment with the proposed relay server and compare its metrics with those of a commercial relay server. In all scenarios, we used a Meta Quest 2 VR HMD networked over a 5GHz Wi-Fi connection.

In the "Softbody" and "MultiObject" scenarios, the PhyS was hosted in an \textit{AMD Ryzen 9 5900X} CPU server with \textit{64GB of 3200MHz }  RAM, and \textit{NVIDIA GeForce RTX2070S} GPU. The PhyS server was connected to a \textit{ZTE H1600} Router via Ethernet cable. 

In the "CCU" scenario, the PhyS was hosted on an Intel Core i7-6700K CPU server with 32GB of 2133MHz RAM and an NVIDIA GeForce GTX1070 GPU. We spawned special bot-users in the XR session that ran on local computers connected via Ethernet cable to the ZTE H1600 router, but connected to the PhyS over the internet using Ethernet.

\begin{figure}[h]
  \centerline{\includegraphics[width=8cm]{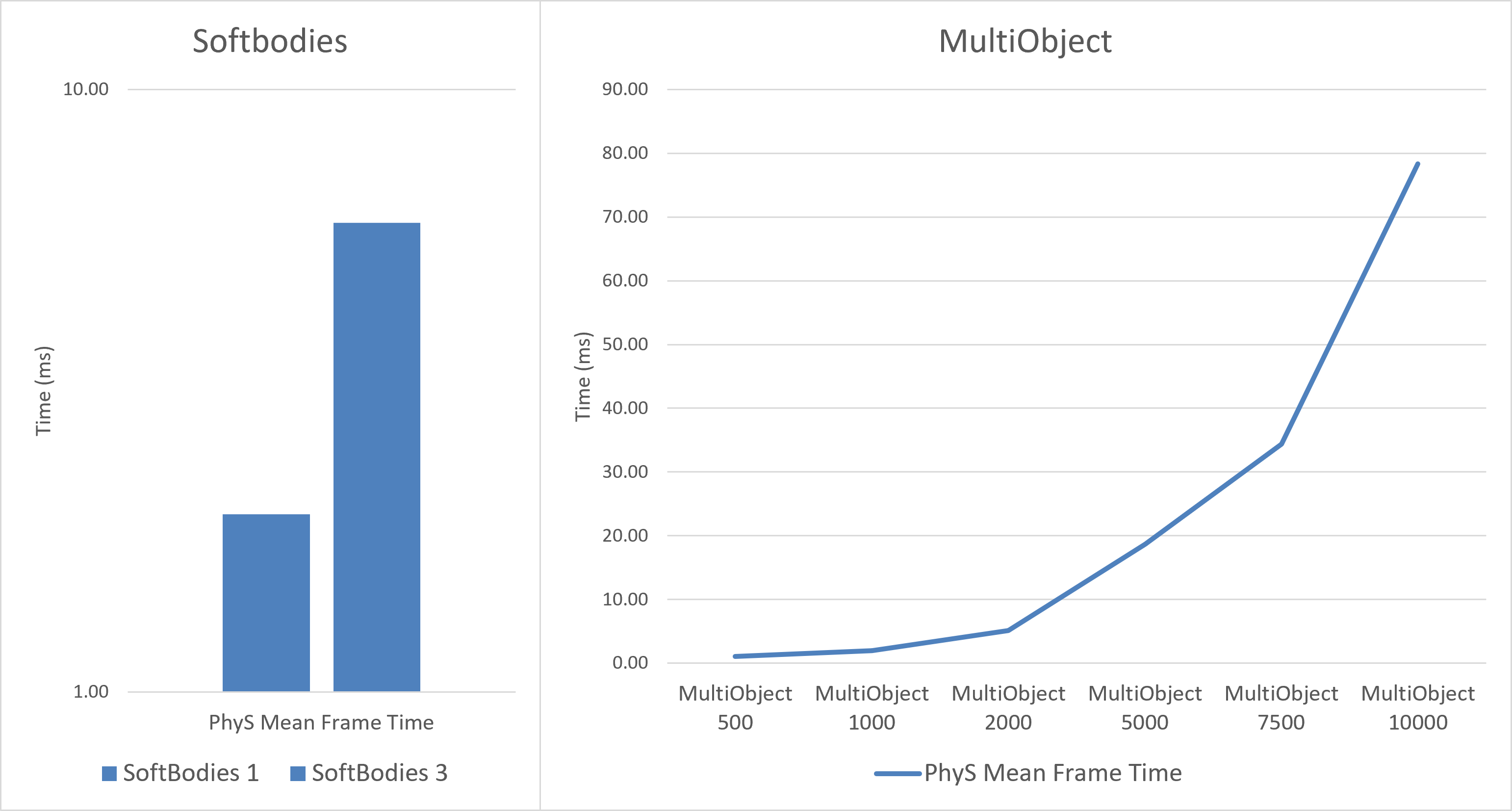}}
\caption{PhyS performance per scenario}
\label{fig:PhySPerformance}
\end{figure}
\begin{figure}[h]
\centerline{
  \includegraphics[width=8cm]{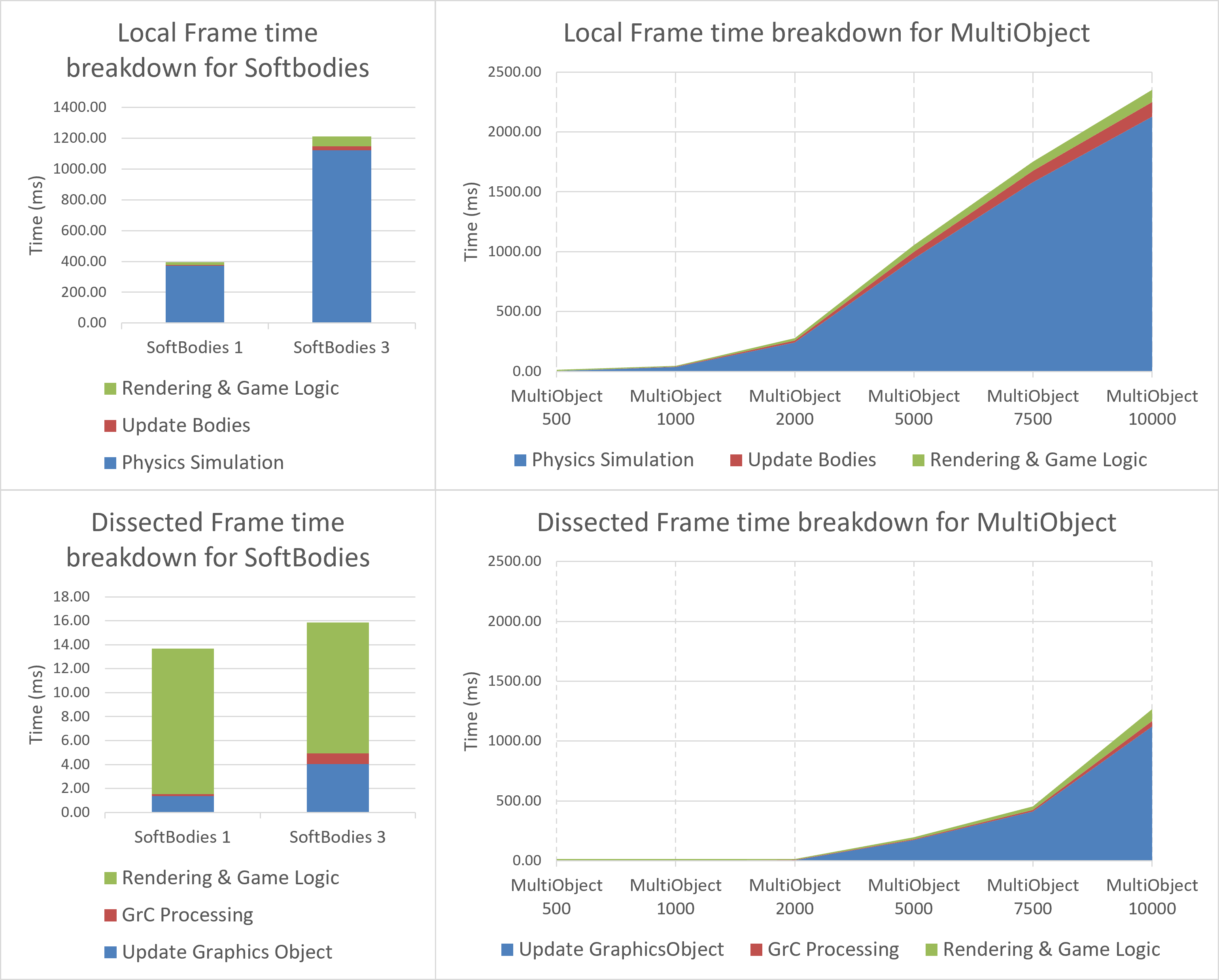}}
\caption{Average frame time breakdown on HMD for local-physics (top) vs decoupled-physics (bottom) computations, for "Softbodies" (left) and "MultiObject" (right) scenarios.}
\label{fig:HMDstacked} 
\end{figure}
\begin{figure}[h]
\centerline{
  \includegraphics[width=8.5cm]{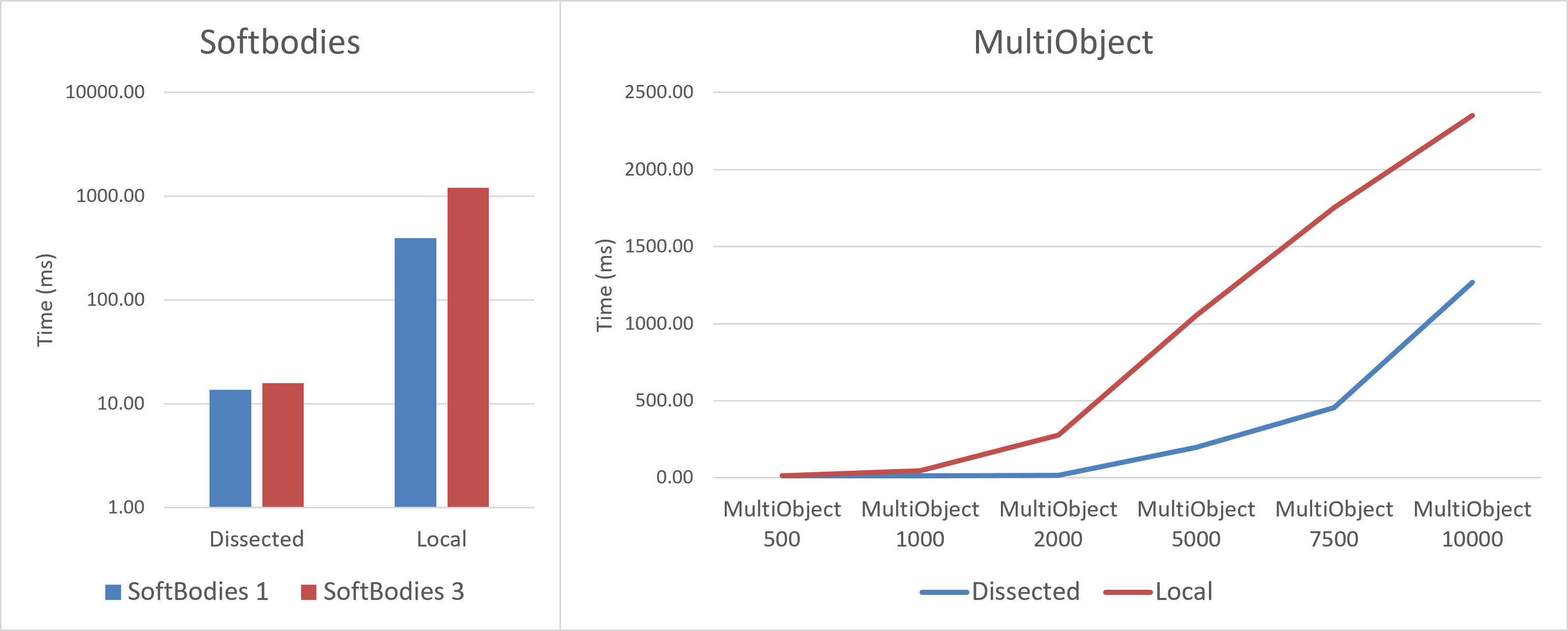}}
\caption{Average total frame times on an HMD for local-physics vs Decoupled-physics, during Softbodies (left) and MultiObject (right) scenarios.}
\label{fig:HMDCurve} 
\end{figure}
\begin{figure}[h]
  \includegraphics[width=8.5cm]{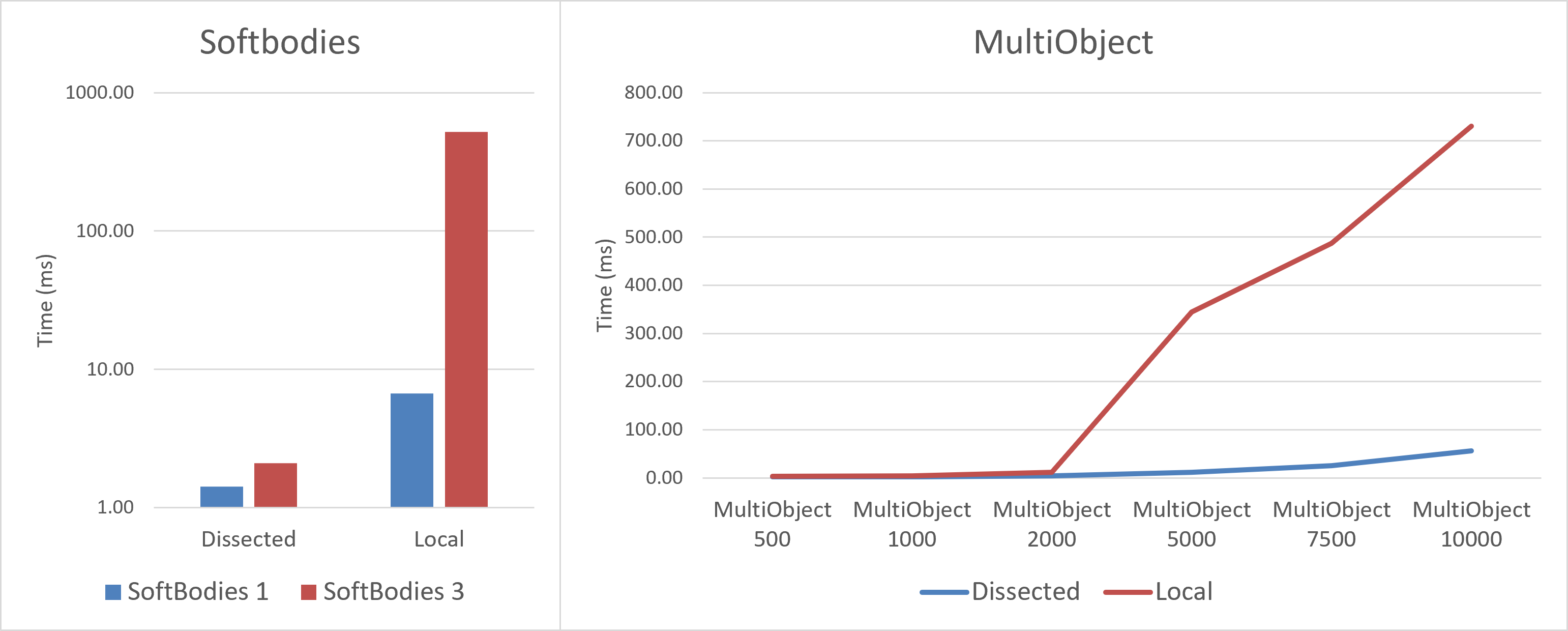}
\caption{Average total frame times on a desktop-PC for local-physics vs Decoupled-physics computations, during Softbodies (left) and MultiObject (right) scenarios.}
\label{fig:PCCurve} 
\end{figure}

MultiObject experimentation showcased that our remote physics approach decreases significantly the total frame time in both HMD (see figure \ref{fig:HMDCurve}) and desktop PC scenarios (see figure \ref{fig:PCCurve}) for all cases. For scenes with more than 7500 objects, we notice that although the Graphics object update process increases significantly for our approach, as the HMD has to perform a great number of updates, it is almost half compared to the local physics case. The breakdown of frame times (see figure \ref{fig:HMDstacked} right) shows that most of the frame time in local-physics case is consumed by physics calculations, which explains the great improvement in the decoupled physics case. The PhyS simulates (see figure \ref{fig:PhySPerformance} right) up to 2000 objects in less than 10ms, while the rest of the cases are far below the local physics case. Additionally, table \ref{tab:Results} shows a steady increase of outgoing throughput from the PhyS as objects increase. 

In the decoupled physics case of the Softbody experimentation on HMD (see figure \ref{fig:HMDstacked} left and figure \ref{fig:HMDCurve}) we see similar values with MultiObject scenario (cases of 500 and 1500 objects), where in the local we notice a great reduction in the total frame time, as all complex softbody computations are handled by the HMD. Although the softbodies have around 500 particles each, they are not equal in performance with 500 objects locally, due to the added computational complexity from spring joints and multiple rigidbodies. In the dissected case, the computational power required from the HMD for softbodies or 500 objects is the same, since GHost just syncs transforms. We also notice (figure \ref{fig:HMDstacked} left) a higher rendering and game logic percentage as the physics offloading allows for more complex rendering without having to worry about the physics overhead. The PhyS performance (see figure \ref{fig:PhySPerformance} shows that both softbodies are simulated in below 10ms time, far below the physics simulations in the local physics case.

In the respective Softbody and MultiObject experimentations with desktop-PC (see figure \ref{fig:PCCurve}) we see much greater improvement in overall performance due to the greater specs in CPU and GPU of the desktop computer. The trends remain the same. Particularly, the 10,000 object scenario is a typical case of this result where the frame time of the dissected physics case is 60ms compared to 750 ms of the local physics case.

\begin{figure}[h]
  \includegraphics[width=8.5cm]{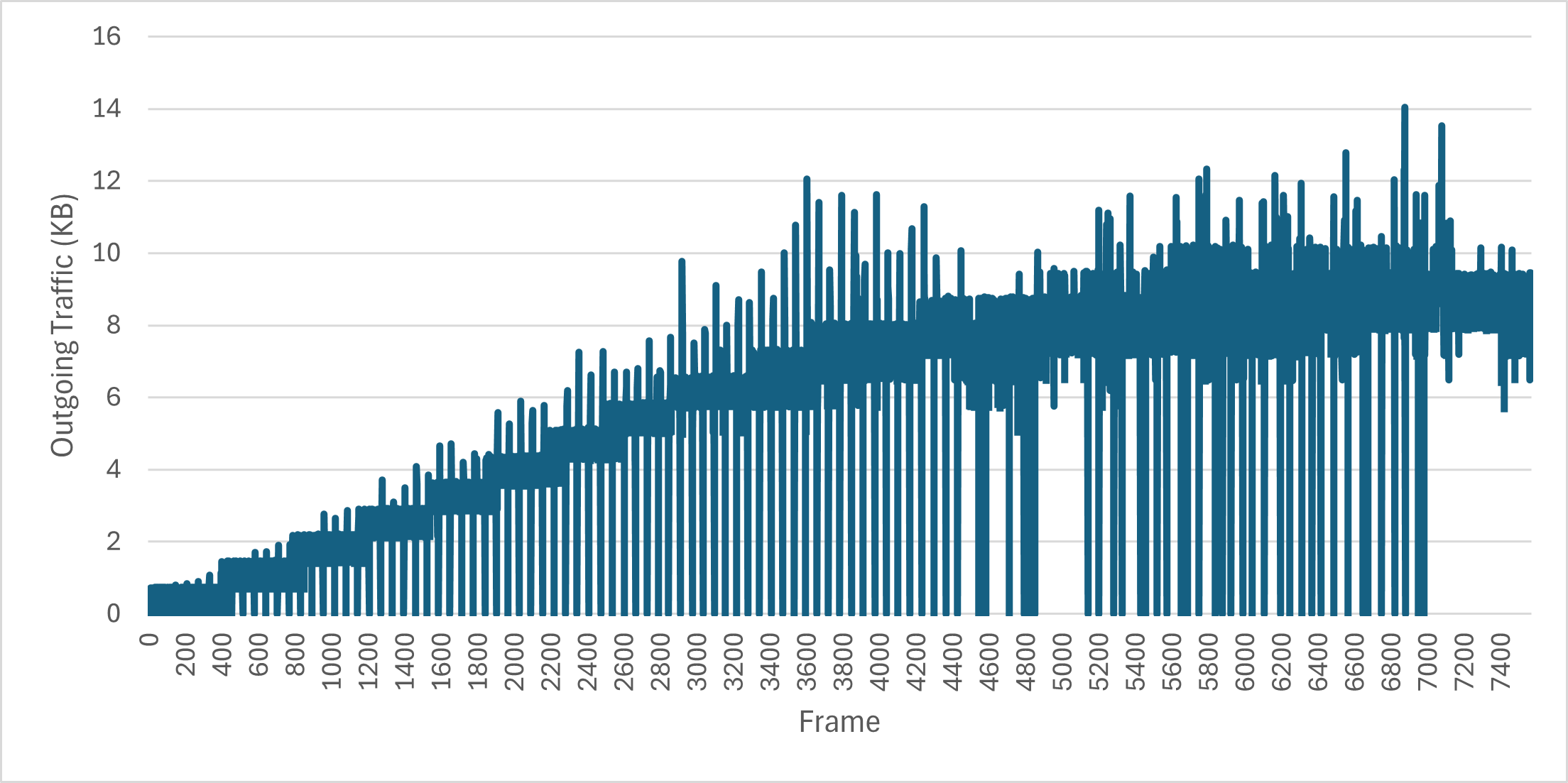}
  \includegraphics[width=8.5cm]{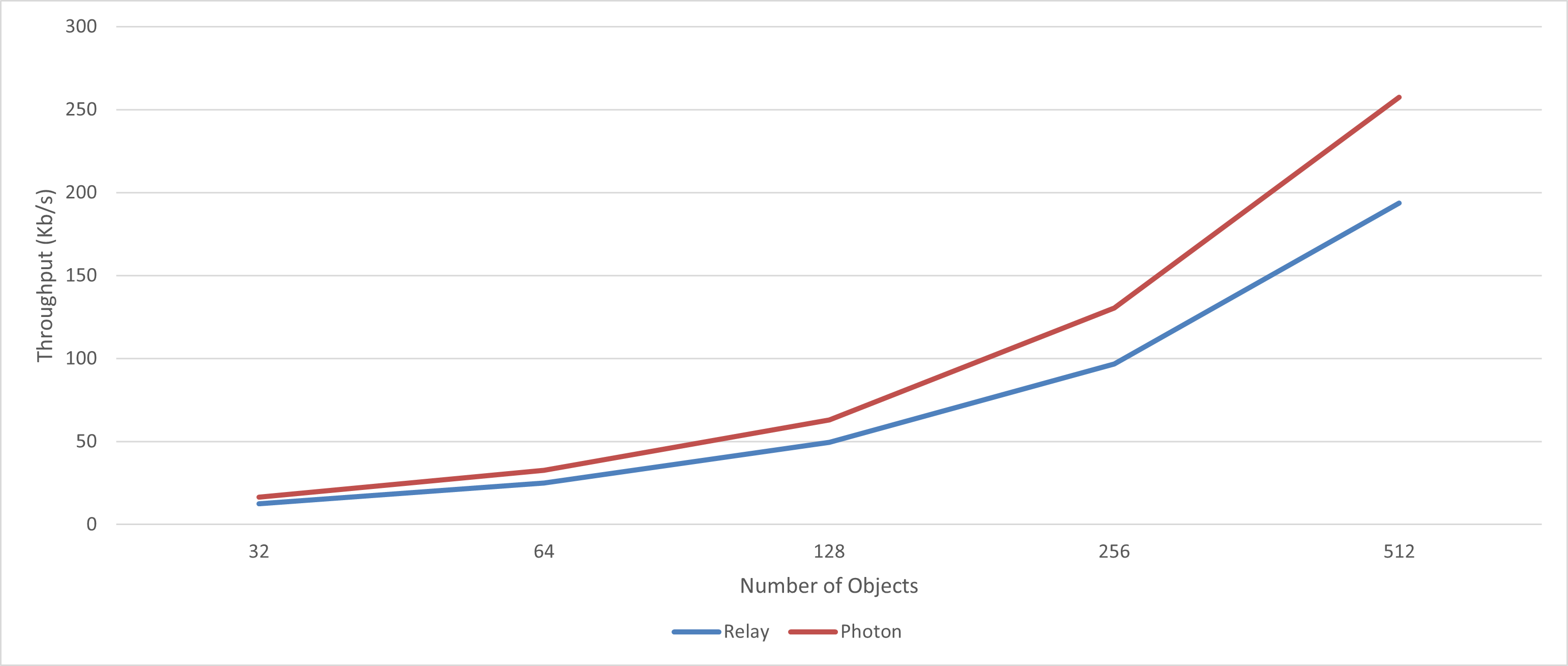}
\caption{Top: PhyS outgoing network usage for the CCU scenario. Bottom: Relay server outgoing bandwidth consumption.}
\label{fig:CCUNet_RelayComp} 
\end{figure}
Experimentation outcomes in the CCU scenario showcases PhyS capability to serve (see figure \ref{fig:ExpCCU_MultiObject}) successfully 100 concurrent users in the same VR session. Due to small number of available XR HMDs, most of the 100 users were bot-users, deployed with same physics objects as an HMD user. Bot-users constantly moving in the scene, so that they generate the same load on the physics server as an HMD user. In that respect, we artificially generated a quite representative standard XR session for the sake of CCU experimentation scenario. 
As users gradually join the session, we notice a steady rise in outbound network usage on the PhyS (see Figure \ref{fig:CCUNet_RelayComp}), which stabilizes after all users have joined the session and are constantly moving in the scene.

\begin{table}
\centering
\caption{Top: Average unreliable outgoing throughput. Bottom: Relay server outgoing bandwidth consumption}
\label{tab:Results}       
\begin{tabular}{ccc}
\begin{tabular}{cr}
\hline\noalign{\smallskip}
\textbf{Experiment} & \textbf{Throughput} \\ 
\noalign{\smallskip}\hline\noalign{\smallskip}
Bunny\_1   & 0.46 Mb/s \\
Bunny\_3   & 1.40 Mb/s \\
Obj\_500   & 0.72 Mb/s \\
Obj\_1000  & 1.32 Mb/s \\
Obj\_2000  & 2.70 Mb/s \\
Obj\_5000  & 5.45 Mb/s \\
Obj\_7500  & 8.31 Mb/s \\
Obj\_10000 & 10.95 Mb/s \\
\noalign{\smallskip}\hline
\end{tabular}
\\
\begin{tabular}{crr}
\hline\noalign{\smallskip}
\textbf{Objects} & \textbf{Our Relay} & \textbf{Photon} \\ 
\noalign{\smallskip}\hline\noalign{\smallskip}
32   & 12.56 KB/s & 16.52 KB/s \\
64   & 24.86 KB/s  & 32.58 KB/s \\
128  & 49.50 KB/s   & 63.00 KB/s \\
256  & 96.80 KB/s   & 130.50 KB/s \\
512  & 193.60 KB/s  & 257.40 KB/s \\
\noalign{\smallskip}\hline
\end{tabular}
\end{tabular}
\end{table}

The total latency in multiuser scenarios refers to the delay experienced by UserA in viewing an interactable object that UserB interacts with. The computation of the worst-case total latency consists of five components: a) the frame rendering time for both UserA and UserB, b) the network latency between the PhyS and both UserA and UserB, and c) the PhyS computation time. To reduce network load, PhyS sends updates for interacted objects at a rate of 48 times per second, introducing a delay of 20ms between updates. A sub-experiment for total latency was conducted with two users in the same session moving a virtual object around. In this experiment, the HMDs were connected to the same Wi-Fi network, while the PhyS was situated remotely and accessed via the Internet. The resulting average total latency was measured to be approximately 68ms (see figure \ref{fig:PhySPerformance}), which provides evidence that the decoupled Physics server can be used without compromising QoE in interactive multiuser scenarios.

Experimentation with the relay server, with cases involving up to 512 physics objects, showcased a constant improvement in outgoing bandwidth compared to the commercial Photon Cloud relay server (free plan) (see Table \ref{tab:Results}). Our experimentation scenario involves multiple objects (cubes), each constantly rotating and translating, to produce extreme cases of network bandwidth per object. Both solutions were set up with a send rate of 12 times per second per object. The host for each session is responsible for sending the latest transform data to the relay server.

In the performance analysis, our relay server appears to be noticeably efficient as it constantly exhibits lower outbound rates, compared to the alternative Photon solution, while the difference between them is constantly increasing (see figure \ref{fig:CCUNet_RelayComp}). 

We assume each user requires four distinct physics objects: two hands, one head (avatar), and one interactable object, which allows us to determine the maximum number of users our relay server can support. Specifically, given a total system capacity of \textit{M} objects, the number of supported users \textit{N} is derived from the formula \textit{N = M/4}. This resource allocation aligns precisely with our relay server's capability to efficiently manage multiple user interactions for up to 128 users.

\section{Conclusions and Future work}\label{sec:6}
We presented a novel system that transforms a modern game engine's pipeline, optimizing XR performances and enhancing user experience in XR environments. By decoupling the physics engine from its tight connection with the game engine pipeline and implementing a client-server N-1 architecture, we establish a scalable solution that efficiently serves multiple graphics clients (HMDs) with a single physics engine application running on edge/cloud infrastructure. This single point of truth for physics computations not only fosters better synchronization in multi-player scenarios, without introducing unnecessary overhead in single-player experiences. The maintenance of the Physics state at the dedicated engine inside the PhyS, regardless of users joining, leaving, or participating in the session, ensures the continuity of the XR session. Additionally, the introduction of a relay server facilitates seamless and optimized collaboration between users utilizing local physics and those connected to the physics server, enabling diverse participation in shared virtual environments. 

The decoupling of the Physics Engine from the HMD and relocating it to an edge/cloud node alleviates strain on local hardware, empowering it to dedicate more resources to rendering high-quality visuals. This strategic offloading of heavy tasks from the CPU unlocks the full potential of untethered HMDs, allowing their powerful GPUs to produce more impressive and complex visuals without being bottlenecked by CPU limitations. 

Moreover, this approach yields numerous benefits including increased frame-rate and QoE in highly interactive and realistic VR simulations, support for advanced interactions on softbodies, scenes with a great number of physics objects, and multi-user scenes with more than 100 CCUs. The modular pipeline with one physics server for all clients ensures efficient resource utilization, facilitates potential gains in HMD battery life and increased user mobility. Also, the higher frame rate achieved translates to less nausea and enables more realistic physics simulations, enhancing the effectiveness of VR medical training applications.

Additionally, the incorporation of a Geometric algebra interpolator minimizes inter-calls between dissected parts, preserving an equivalent QoE while alleviating network stress. Collectively, these design decisions contribute cohesively to the successful achievement of the stated goals, showcasing a well-thought-out and effective approach to optimizing XR gaming experiences.

We demonstrated the feasibility of Physics engine offloading through experimental validation with 100 concurrent users, 10,000 objects and softbody simulations. This capability not only confirms the technical viability of the proposed architecture but also opens up new avenues for enhancing XR experiences, promising more immersive and collaborative XR applications without compromising HMD or desktop performance.

In our future work, we envision to optimize PhyS by enhancing and evaluating QoE even under degraded network conditions. To fully harness edge-cloud resources, we plan to integrate multi-threaded physics algorithms and GPU compute shaders, aiming to significantly reduce physics computation times within PhyS. Exploring Pixar's USD physics schemas presents an exciting opportunity to streamline the initiation process of the decoupled PhyS. Additionally, to further enhance frame times on XR HMDs, we will investigate methods to optimize the update process of Graphics objects. Finally, expanding PhyS functionality to support persistent always-on sessions will also be a key focus, aiming to allow seamless and uninterrupted user experiences across extended periods of interaction in Metaverse.

\section*{Acknowledgments}
This work was partially funded by EU research and innovation programmes CHARITY (H2020 GA No 101016509), FIDAL (Horizon Europe GA No 101096146) and the Innovation project Swiss Accelerator supported by Innosuisse. We would like to thank Manos Kamarianakis and Maria Pateraki for their valuable comments.


\bibliographystyle{plain} 
\bibliography{main-WORKING}       

\newpage

\section*{Author Biography}

\begin{wrapfigure}{l}{25mm} 
\includegraphics[width=1in,height=1.25in,clip,keepaspectratio]{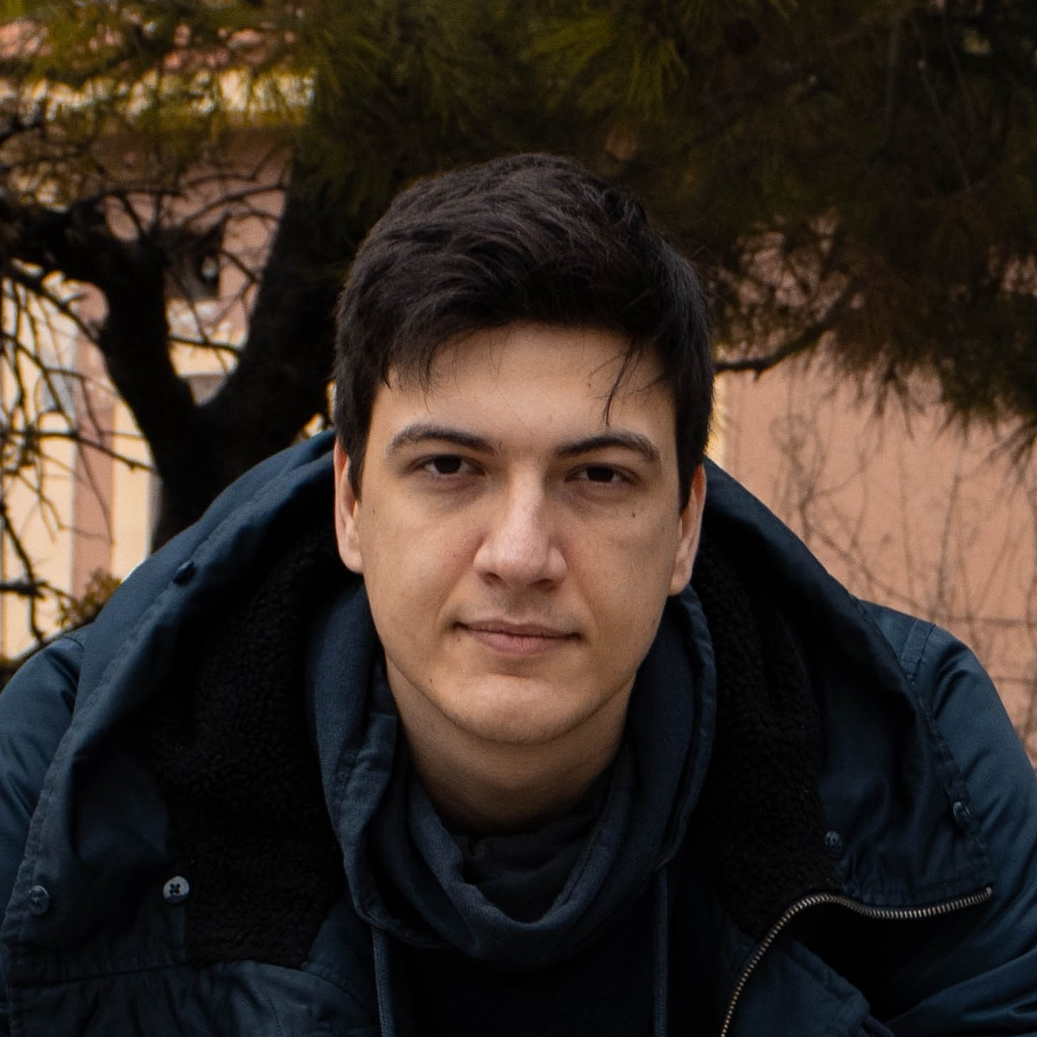}
\end{wrapfigure}
\textbf{George Kokiadis} is PhD Candidate at University of Crete, and a member of the Human-Computer Interaction Lab at FORTH-Hellas. His research revolves around the use of 5G and Cloud Infrastructure to enhance XR Technologies and Applications. Contact him at george.kokiadis@oramavr.com.
\\

\begin{wrapfigure}{l}{25mm} 
\includegraphics[width=1in,height=1.25in,clip,keepaspectratio]{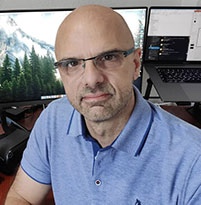}
\end{wrapfigure}\par
\textbf{Dr. Antonis Protopsaltis} is a Computer Scientist and the Lead Research Scientist at ORamaVR. He is a Special Teaching Fellow in Computer Graphics at the University of Western Macedonia (UoWM) and an Affiliated Researcher at the ITHACA-UOWM lab, specializing in Extended Reality and CAD methods. Contact him at antonis.protopsaltis@oramavr.com.
\\

\par
\begin{wrapfigure}{l}{25mm} 
\includegraphics[width=1in,height=1.25in,clip,keepaspectratio]{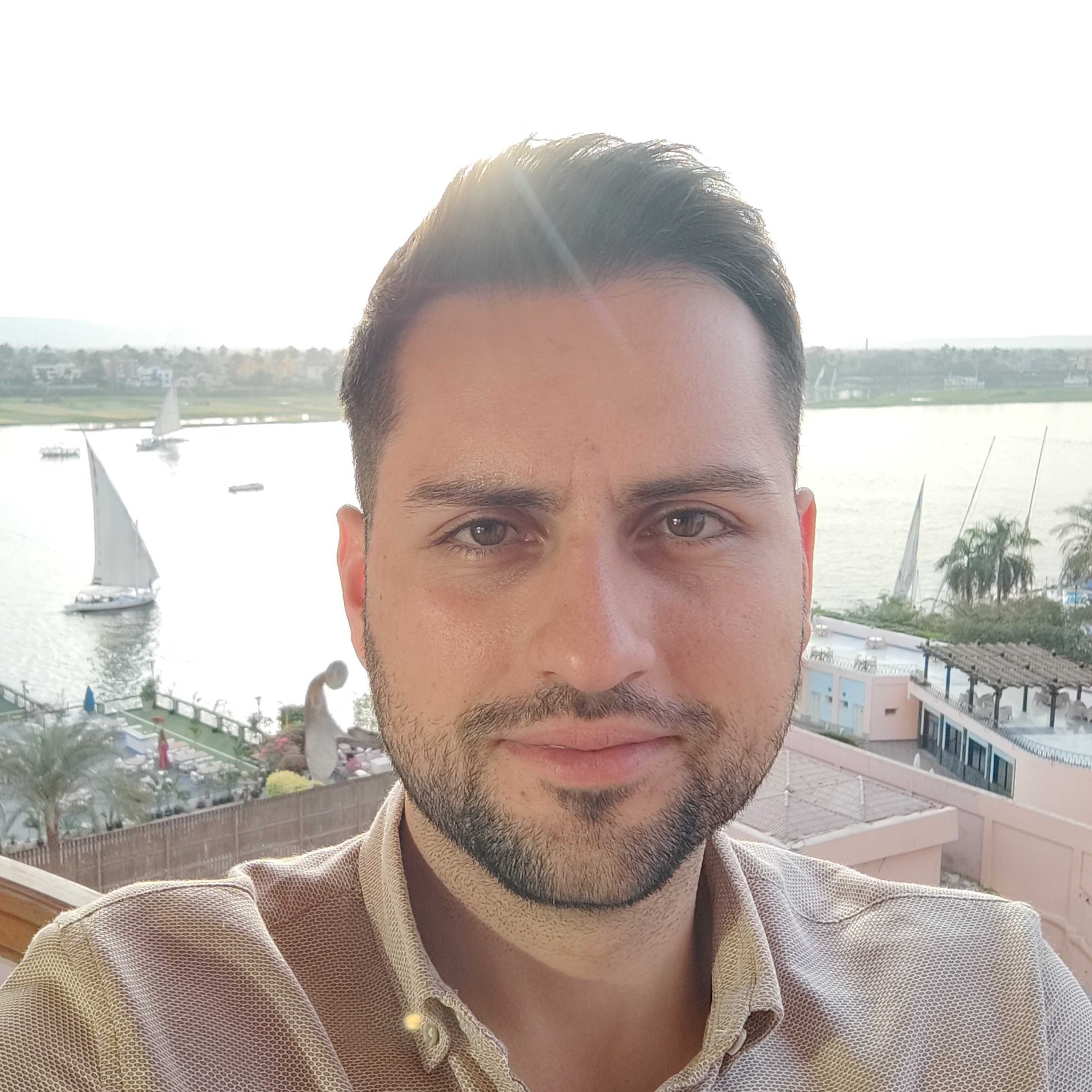}
\end{wrapfigure}
\textbf{Michalis Morfiadakis} is a MSc student at the Computer Science Department of the University of Crete, and  a Networking Developer at ORamaVR. His thesis was about developing a Relay Server-based Network Solution for VR Collaborative Applications. Contact him at michael.morfiadakis@oramavr.com.
\\

\begin{wrapfigure}{l}{25mm} 
\includegraphics[width=1in,height=1.25in,clip,keepaspectratio]{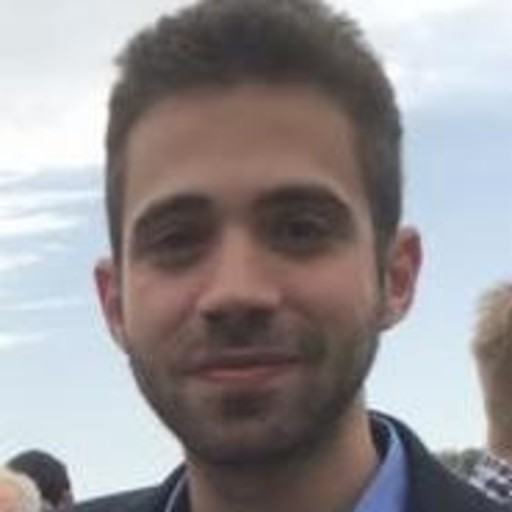}
\end{wrapfigure}\par
\textbf{Nick Lydatakis} is Head of platform  at ORamaVR and a PhD Candidate at University of Crete, 
exploring application partitioning frameworks for high-fidelity edge-cloud collaboration in Extended Reality with soft mesh deformations. Contact him at nick.lydatakis@oramavr.com.
\\

\begin{wrapfigure}{l}{25mm} 
\includegraphics[width=1in,height=1.00in,clip]{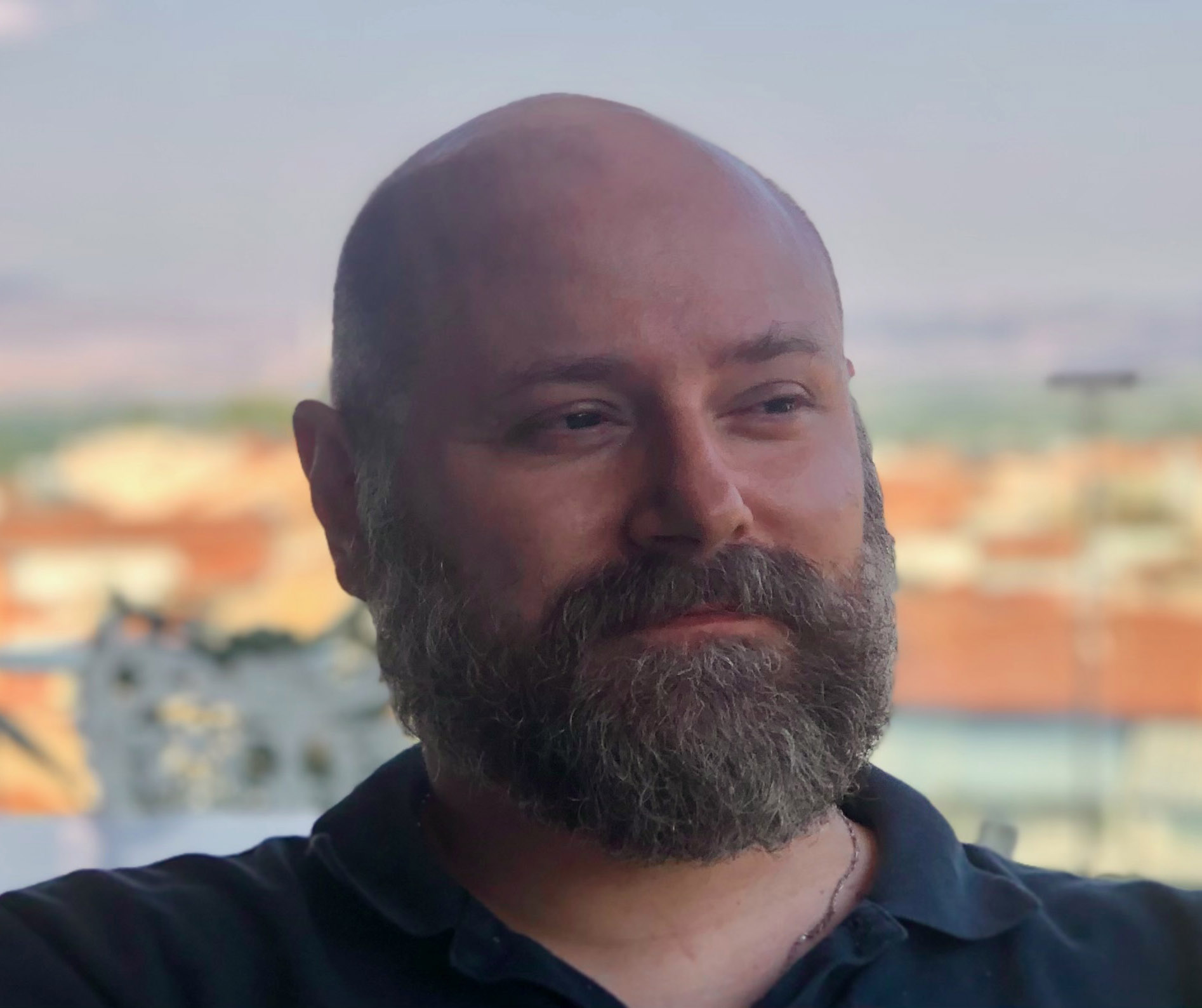}
\end{wrapfigure}\par
\textbf{Dr. George Papagiannakis} is a distinguished computer scientist with a specialization in computer graphics, extended reality, and geometric algebra. After earning his PhD from the University of Geneva in 2006, he has established a notable academic and entrepreneurial career. Currently, he is a professor at the University of Crete and holds positions at FORTH-ICS and the University of Geneva. He has made significant contributions to human-computer interaction and virtual reality, particularly in medical training and virtual heritage, using advanced computational and geometric computer graphics models.
\noindent
As a co-founder and CEO of ORamaVR, he pioneers in developing VR solutions for medical training. He has published over 120 papers and is involved in various professional societies including IEEE and ACM. Notably, his work is recognized through awards like the Marie-Curie Fellowship and he has held key roles in prominent conferences such as CGI. His book on Mixed Reality and Gamification has been highly influential, evidencing his impact in the field. Dr. Papagiannakis continues to lead research projects, supervising numerous doctoral and master’s students, and has attracted significant R\&D funding. Contact him at george.papagiannakis@oramavr.com.

\end{document}